\documentclass{osa-article}

\journal{oe}

\articletype{Research Article}

\usepackage{array}
\usepackage{comment}
\usepackage{bbold}

\begin{document}

\title{Physics-based neural network for non-invasive control of coherent light in scattering media}

\author{Alexandra d'Arco,\authormark{1} Fei Xia,\authormark{1} Antoine Boniface \authormark{1,2}, Jonathan Dong\authormark{1,3} and Sylvain Gigan\authormark{1,*}}

\address{\authormark{1}Laboratoire Kastler Brossel, ENS–Universite PSL, CNRS, Sorbonne Université, Collège de France, 24 Rue Lhomond, F-75005 Paris, France.\\
\authormark{2}Laboratory of Applied Photonics Devices, École Polytechnique Fédérale de Lausanne (EPFL), CH-1015 Lausanne, Switzerland\\
\authormark{3}Biomedical Imaging Group,  École Polytechnique Fédérale de Lausanne (EPFL), CH-1015 Lausanne, Switzerland}

\email{\authormark{*}sylvain.gigan@lkb.ens.fr} 



\begin{abstract} Optical imaging through complex media, such as biological tissues or fog, is challenging due to light  scattering. 
In the multiple scattering regime, wavefront shaping provides an effective method to retrieve information; it relies on measuring how the propagation of different optical wavefronts are impacted by scattering. Based on this principle, several wavefront shaping techniques were successfully developed, but most of them are highly invasive and limited to proof-of-principle experiments. Here, we propose to use a neural network approach to non-invasively characterize and control light scattering inside the medium and also to  retrieve information of hidden objects buried within it. Unlike most of the recently-proposed approaches, the architecture of our neural network with its layers, connected nodes and activation functions has a true physical meaning as it mimics the propagation of light in our optical system. It is trained with an experimentally-measured input/output dataset built from a series of incident light patterns and corresponding camera snapshots. We apply our physics-based neural network to a fluorescence microscope in epi-configuration and demonstrate its performance through numerical simulations and experiments. 
This flexible method can include physical priors and we show that it can be applied to other systems as, for example, non-linear or coherent contrast mechanisms.
\end{abstract}

\section{Introduction}

Coherent light is subject to scattering when it propagates through optically heterogeneous samples such as biological tissues. Ballistic light is exponentially attenuated with depth, and the remaining light is scattered giving rise to a complex interference pattern, also called speckle \cite{goodman1976some}. Most imaging techniques rely on the use of ballistic light and are therefore limited to shallow depths. In the past few years, many advances have been made for imaging at depth, mostly by controlling the phase and/or amplitude of the light being scattered thanks to spatial light modulators (SLMs) \cite{Mosk2012SLM,Rotter2017TMformalism}. 

The first wavefront shaping experiment consisted in iteratively optimizing the SLM phase pattern using feedback metrics, such as signal strength, to focus the transmitted scattered light onto a diffraction-limited spot \cite{Vellekoop2007}. This was a major step forward for optical imaging in turbid medium as scanning the spot across the sample offers an image. Later on, the transmission matrix (TM) that characterizes the linear and deterministic propagation of the optical field through the medium has been experimentally measured \cite{Ishimaru1997WavePA,Popoff2011TM}. Knowing the TM gives more information than optimizing the input field and can also be used for imaging \cite{popoff2010imaging}. However in both cases, a feedback signal from the focal plane is needed. A detector (a single pixel detector or a camera) must then have a direct access to the focal plane which is in most scenarios highly invasive. This problem can be overcome by using guidestars, such as nonlinear signals, to provide feedback in a non-invasive way \cite{Horstmeyer2015Guidestar,Chaigne2014photoacoustic}, but some contrast mechanisms, such as linear fluorescence, are left aside as not easy to use. A few linear fluorescence settings were proposed for imaging but are only able to reconstruct the superficial layers of biological samples with high resolution \cite{lichtman2005fluorescence,webb2012epi,ghielmetti2014direct,hofer2018wide}. Lately, a few approaches have been developed for non-invasive imaging in scattering media with fluorescence feedback, but they are specific and not very flexible \cite{Dayan2020noninv, Bertolotti2012noninv}.  

Meanwhile, several techniques that utilize emerging computational tools, such as neural networks, have emerged in parallel for controlling light through scattering media such as diffusers and multimode fibres \cite{Faccio2019MMF,Barbastathis2019Optica_DL}. Although these methods offer the advantage of being physics-informed, they are primarily used in transmission and remain invasive. Moreover, neural networks approaches that are non invasive often include several layers, making it more difficult to interpret the neural network physics and thus add prior physical knowledge \cite{Turpin2018NN}.

To overcome these issues, we propose a versatile neural network-based TM retrieval method for non-invasive light focusing through scattering media \cite{Mounaix2018scattmed}. The approach is based on mapping the input/output information in a microscopy configuration thanks to a 2-layer neural network. It is simple, model-based, and applicable to a variety of imaging scenarios. We present experiments on non-invasive fluorescence imaging and simulations generalizing this concept to a different contrast mechanism (Second Harmonic Generation, SHG).  

\section{Principle}
\begin{figure}[!ht]
\centering
\includegraphics[width=0.5\textwidth]{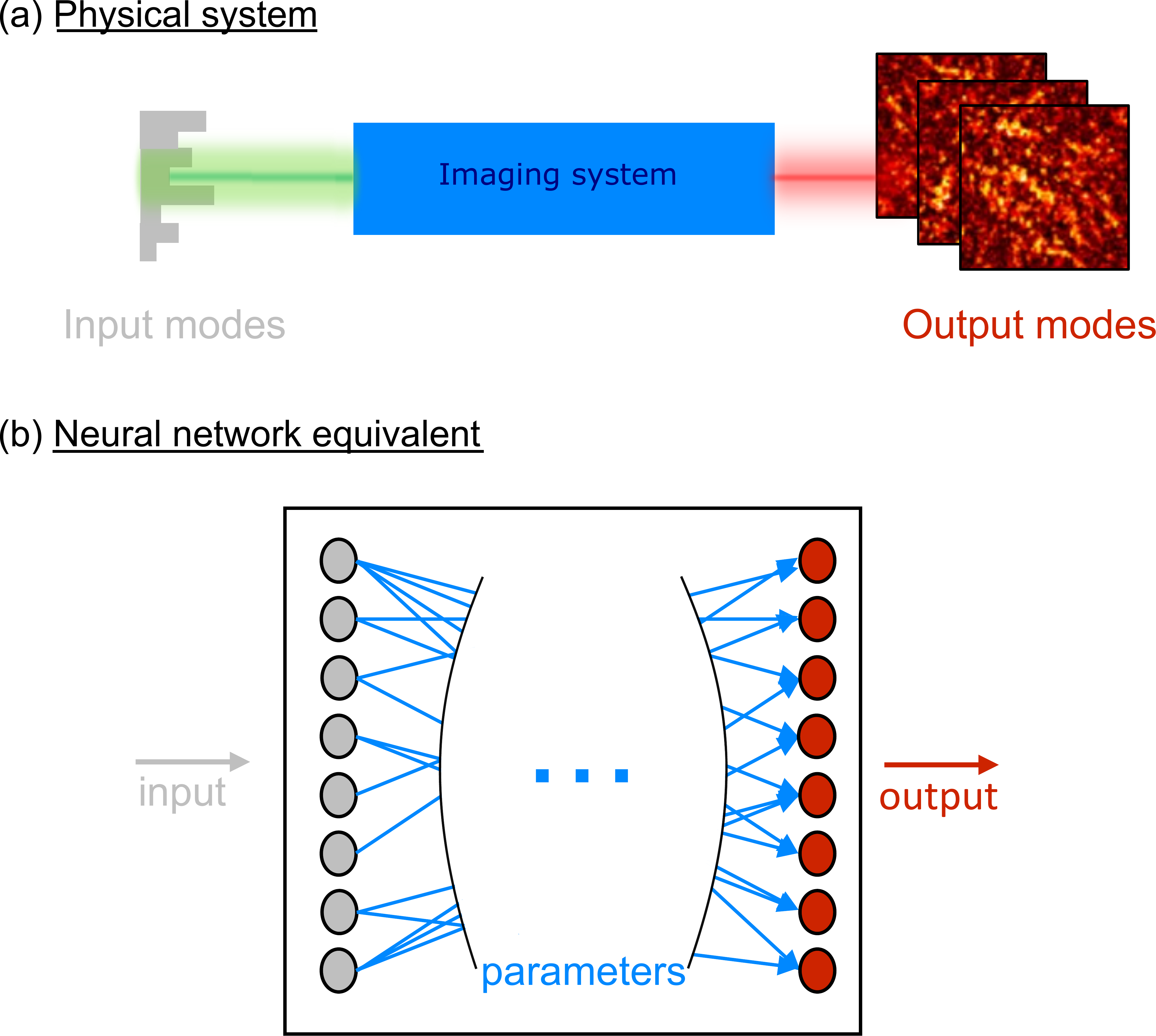}
\caption{\textbf{Neural network representation of a physical system.} (a) Physical system : from input and the features of the optical system (blue box), output modes such as images can be obtained. (b) Neural network representation of this physical system. The weight of the connection are the physical system parameters.}
\label{figure_0}
\end{figure}

In the general context of non-invasive imaging in scattering media, the propagation of light can be decomposed into two steps: the forward path (from the light source to the object of interest, corresponding to the illumination) and the backward path (from the object to the detector, corresponding to the signal re-emitted by the object). Due to scattering, the light waves in both paths are strongly distorted and as such no information can be retrieved directly. Inspired from the classical experiment to measure the TM, our strategy is to introduce an SLM to modulate the incident wavefront, and simply measure the corresponding re-emitted signal from the object with a camera. Most importantly, the latter captures the light that has traveled both the forward and backward paths, unlike in a conventional TM experiment where only the forward path is characterized. As this non-invasive geometry is more complicated to describe in terms of light propagation, it can be seen as a black box describing the relation between the input (on the SLM) and output (on the camera) patterns. It is then possible to make a "digital twin" of the system by describing this black box with a multi-layer neural network with the right constraints \cite{grieves2019virtually}. 
Conventional neural networks are built from an operational layer that involves a trainable matrix-vector multiplication followed by an element-wise nonlinear activation. The weights of the matrix-vector multiplication are adjusted during training in order for the neural network to implement a desired mathematical operation \cite{wright2021deep}. In the digital twin approach, the operation is linked to the physical properties of the system. By changing the network weights, one can alter the physical transformation performed on the input data in order to learn these parameters according to the physical system. Our approach is based on this representation (see Fig.~\ref{figure_0}).

\begin{figure}[!ht]
\centering
\includegraphics[width=1.0\textwidth]{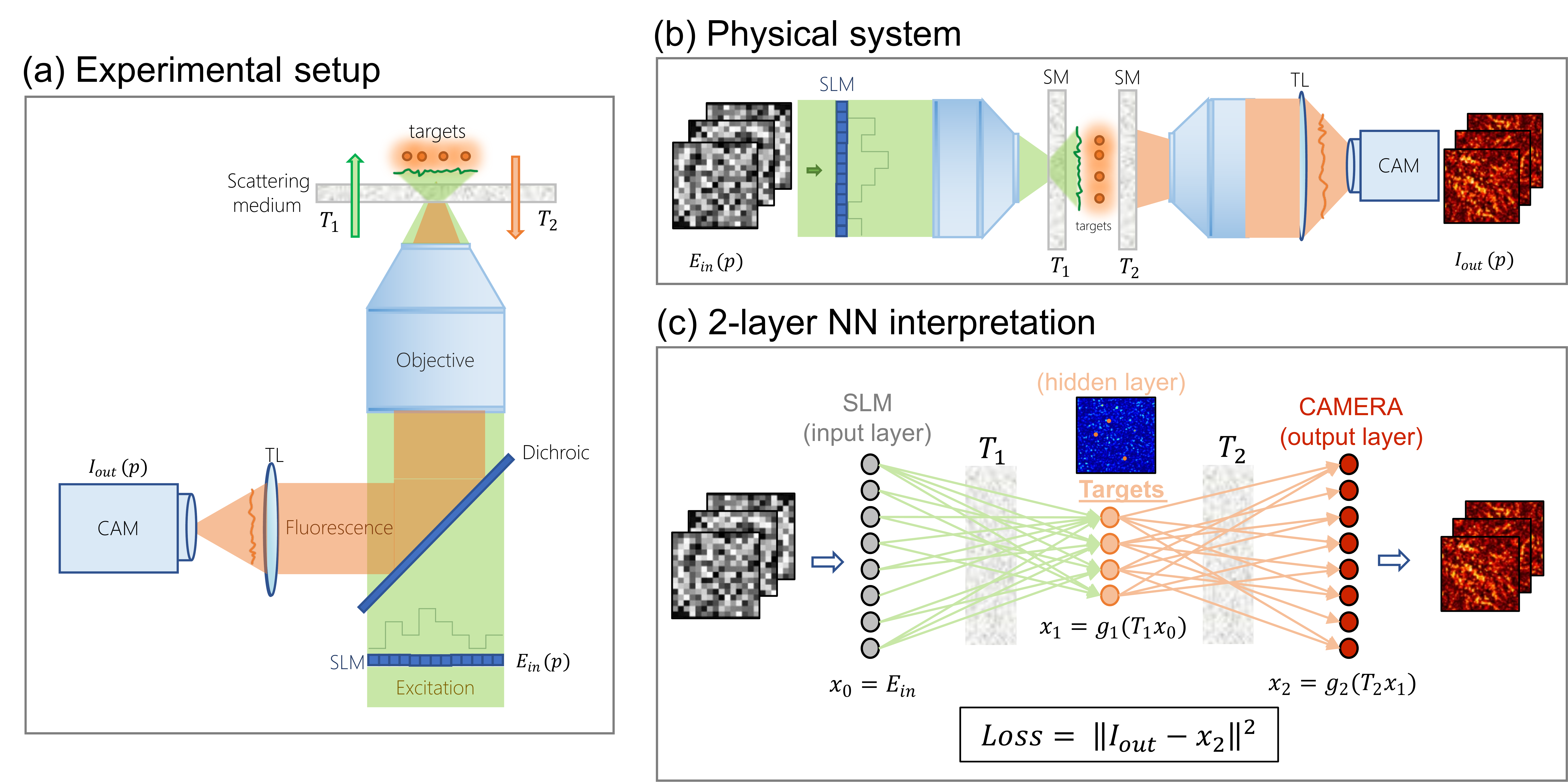}
\caption{\textbf{Modeling fluorescence imaging setup as a Neural Network.} (a) Schematic view of the non-invasive experimental setup. (b) Schematic view of the unfolded experimental setup. A randomly modulated speckle pattern illuminates a fluorescent object (beads or pollen seeds), hidden behind the scattering medium. $E^{\text{exc}} = T_1 E^{\text{in}}$ excite the object, which emits a fluorescence signal in return, that is backscattered by the medium and detected in epi-geometry on a camera : $I^{\text{out}} = T_2 |E^{\text{exc}}|^2 $. TL : tube lens. (c) Neural network mimicking the experimental setup. The neural network is trained to regenerate fluorescence speckle $x_2 \approx I^{\text{out}}$ from the input pattern $E^{\text{in}}$ displayed onto the SLM. $x_2$ is compared with the actual original image $I^{\text{out}}$ (record in reflection on a camera) through the $Loss = ||I^{\text{out}} - x_2 ||^2$ function. Using gradient descent, back-propagation updates the weights ($T_1$ and $T_2$) to minimize the $Loss$}
\label{setup}
\end{figure}

We apply this concept to a microscopy imaging system. The physical system and its neural network interpretation are depicted in Fig.~\ref{setup}.
The weights of the network connect all inputs to outputs of the system (from SLM to camera) through a fully connected two-layer network. The weights of each layer physically correspond to the coefficients of matrices $T_{1}$ and $T_{2}$ as we can see in Fig.~\ref{setup}(b). 
The first layer connects the input field $E^{\text{in}}$ to the excitation field $E^{\text{exc}}$ at the object plane. The weights of this layer correspond to the complex-valued coefficients of the matrix $T_{1}$. The size of this matrix is $N_{\text{target}} \times N_{\text{SLM}}$, where $N_{\text{target}}$ is the number of emitters and $N_{\text{SLM}}$ the dimension of SLM patterns. The weights of the second layer represent $T_{2}$, of size $N_{\text{CAM}} \times N_{\text{target}}$ with $N_{\text{CAM}}$ the number of camera pixels. According to the physical nature of the process, 
the forward model can be written as:
\begin{equation}
    \begin{centering}
    x_2 = g_2(T_{2}g_1(T_{1}x_0))
    \end{centering}
\end{equation}
where $x_0 = E^{\text{in}}$ is the input layer, matrices $T_1$ and $T_2$ constitute the weights of the networks and $g_1$ and $g_2$ are the activation functions, defined according to the imaging process \cite{sharma2017activation}. $g_1$ and $g_2$ are physical priors, they are chosen to mimic our optical system. 

In order to find the weights, the network is trained with a dataset composed of $p$ input-output pairs \{$E^{\text{in}}$, $I^{\text{out}}$\}. The gradient of the $Loss$ function is computed, with respect to the weights of the network (i.e. back propagation \cite{lecun1988theoretical}). As it is appropriate for regression predictive modeling problems, the loss function we used in the work is simply a $L_2$-norm loss:
\begin{equation}
    \begin{centering}
    Loss = ||{I^{\text{out}}-x_2}||^2
    \end{centering}
\end{equation}

After a given number of iteration corresponding to a significant decrease of the loss, the weights of the connection are trained. As a result, these weights correspond to the experimental transmission matrices of the system, $T_1$ and $T_2$. This result is experimentally validated by focusing on the fluorescence beads using the retrieved $T_1$ or reconstructing the image using $T_2$. Selective and non-invasive focusing onto beads is done by phase conjugating the retrieved $T_1$ \cite{fisher2012optical,yaqoob2008optical}. If the focus can be obtained onto most beads, the network learned the physical transform at stake in the system with a given accuracy. 

To validate the method, we implement it experimentally in a fluorescence imaging scenario also and also in simulation with a different contrast mechanism (SHG). A different contrast mechanism can be studied by simply modifying the non-linear activation functions of the network, without changing the overall 2-layer architecture. This shows the simplicity of our network and its resulting flexibility since it can be easily applied to other non-invasive model and non-linear imaging settings. 

The experimental setup is shown in Fig.~\ref{setup}. A sparse fluorescent sample made of $1$~\textmu m beads is placed behind a scattering medium. In order for the 2-layer model to reconstruct the object, all beads are placed in the same plane. An SLM is used to modulate the phase of the incident optical field in order to send different illumination patterns to the sample. Experimentally, $p$ random phase masks are displayed on the SLM, denoted $E^{\text{in}}(p)$. The phase of the $N_{\text{SLM}}$ independent pixels is randomly chosen between $0$ and $2\pi$, following a uniform distribution. After its propagation through the scattering medium (which corresponds to a first matrix multiplication), a random speckle field $E^{\text{exc}}(p)$ is formed and illuminates the fluorescent object. During this first step, the transformation of the optical field between the SLM and the sample plane is linear and deterministic. It is represented by the matrix $T_{1}$ that maps the SLM input optical field $E^{\text{in}}$ to the field on the sample in a linear way $E^{\text{exc}} = T_1 E^{\text{in}}$. The activation function of the first layer is then $g_1 = |\hspace{0.5mm}.\hspace{0.5mm}|^2 $, corresponding to the fact that fluorescence signal is proportional to the intensity of the excitation.  In the second step, the fluorescence signal is transmitted through the scattering medium and collected by the camera, a process that can be described by an incoherent and positive matrix $T_2$: as fluorescence is spatially incoherent, the captured camera image is the sum of the fluorescence speckles emitted by each fluorescent source : $I^{\text{out}} = T_2 |E^{\text{exc}}|^2$. In order to fit our physical system we ensure $g_2$ to be the activation function of the second layer defined by $g_2(x) = (x)$, the identity function.
We define here \textit{eigen patterns}, the speckles that are generated by each individual fluorescent emitter (of the focal spot size) on the camera. We made the assumption that the system (scattering medium and fluorescent sample) is static, therefore the \textit{eigen patterns} are the same over time. Thus, the fluorescence signal $I^{\text{out}}(p)$ can be written as:
\begin{equation}
I^{\text{out}}(p) = T_2|E^{\text{exc}}(p)|^2 = T_2|T_1 E^{\text{in}}(p)|^2.
\end{equation}

The experiment is completed by two sets of simulations, one to study our linear fluorescence setup, and another to simulate a different contrast mechanisms (SHG) to show how the activation functions and initial conditions on the TMs can be adjusted. Here, a numerical dataset consisting of pairs of input patterns displayed on the SLM, $E^{\text{in}}$, and corresponding speckle patterns $I^{\text{out}} = g_2(T_2g_1(E^{\text{in}}))$ are generated. The values of $E^{\text{in}}$ are passed to the two fully connected layers which is equivalent to multiplications by the approximate matrices $\widetilde{T_1}$, $\widetilde{T_2}$. 
An approximate image is then obtained as $x_2 = g_2( \widetilde{T_2} g_1(\widetilde{T_1}x_0))$. Initially, $\widetilde{T_1}$ and $\widetilde{T_2}$ are random matrices, complex and real positive-valued, respectively. The derivatives of the $Loss$ function, with respect to the elements of $\widetilde{T_1}$ and $\widetilde{T_2}$ are computed. A stochastic gradient descent approach (SGD or Adam optimizer in Pytorch) is then applied to gradually update the estimated matrices such that it effectively reduces the overall loss function, and the process is repeated for a fixed number of iterations or \emph{epochs}, ensuring convergence of the loss function to a minimum value \cite{bottou1991stochastic}. In this simulation, no noise is added neither on the input nor output data. We also notice that the noise does not significantly impact the reconstruction as long as enough training samples in the dataset can be obtained. (see Supplementary information - Noise Influence to see the impact of noise on the reconstruction). 

\section{Results}

\subsection{Simulations} 

\subsubsection{Discrete object} 
\begin{figure}[!ht]
\centering
\includegraphics[width=1.0\textwidth]{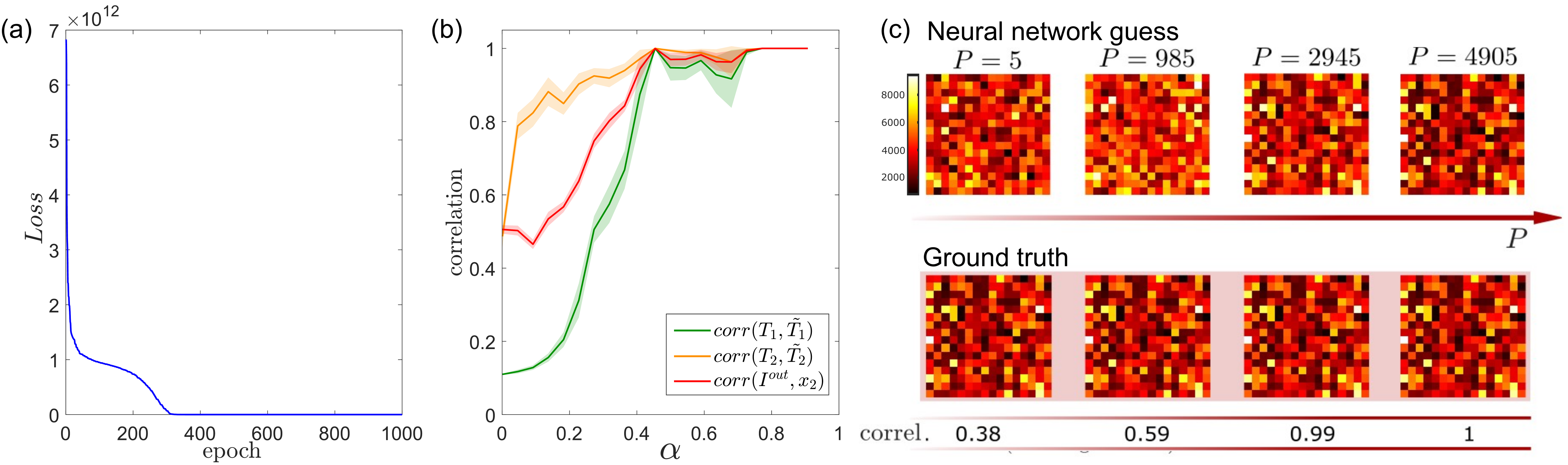}
\caption{\textbf{Simulation results of transmission matrices retrieval.} (a) Loss decrease after 1000 epoch of gradient descent for a given training set size. (b) Correlation between the ground truth $I^{\text{out}}$ and the neural network guess $I^{\text{test}}$, and the ground truth $T_1$ and the neural network guess $\widetilde{T_1}$, and finally between the ground truth $T_2$ and the neural network guess $\widetilde{T_1}$, according to $\alpha = P/N_{\text{pat}}$, with P the training set size. Since the procedure is repeated 10 times for 10 different datasets, the plots represent the mean correlation and its standard deviation in shade. (c) Evolution of the reconstruction of the output image of the test set according to the training set size.}
\label{simu_discrete}
\end{figure}

As a first step, we consider a linear fluorescence forward model based on the two-layer neural network architecture and a discrete object.  
In machine learning, a test set is typically used to evaluate the fit of a model on a training set \cite{mitchell2010traintest}.
The test set is a part of the original dataset which is set aside and used afterwards to assess the performance of the neural network.
We choose a discrete fluorescent object composed of $N_{\text{target}} = 8$ targets, and set the dimension of the hidden layer to correspond to the number of targets. The dimensions of input layer and output layer are $N_{\text{SLM}} = 256$ and $N_{\text{CAM}} = 256$ respectively. Ground truths $T_1$ and $T_2$ are randomly generated following a standard normal distribution, we assume no correlation and no noise is added in the generated ground truth data. We use a training set up to $N_{\text{pat}}=$4900 examples to estimate the weights of the matrices $\widetilde{T_1}$ and $\widetilde{T_2}$.
Then a test set of 500 input examples on the SLM, {$E^{\text{test}}$} (unseen previously by the neural network) is passed to the trained network which generates 500 output images $I^{\text{test}}$. Correlations between $I^{\text{test}}$ and $I^{\text{out}}$; $T_1$ and $\widetilde{T_1}$ ; $T_2$ and $\widetilde{T_2}$ are plotted with respect to $\alpha = P/N_{\text{pat}}$ on Fig. \ref{simu_discrete}, where $P$ is the training set size. This procedure is averaged over 10 repetitions for 10 different data sets (i.e different $T_1$ and $T_2$). See supplementary information for the choice of the rank, i.e. the size of the hidden layer. 

As can be seen in Fig. \ref{simu_discrete}, the loss function is minimized as the number of epochs grows. The correlations between ground truths and neural network guesses of fluorescent output speckle patterns reach a value close to 1 when the training set size increases, meaning that the transmission matrices were well retrieved by the training of the 2-layer neural network. 
Further, we visualize the reconstruction of both matrices, and observe that the dynamics are different based upon the size of the training set. With small training set sizes, $P<500, \alpha < 0.09$, it seems that the best way to minimize the global loss function is to adjust the coefficients of $\widetilde{T_2}$. When backpropagating the gradient, it rapidly converges towards $T_2$ (correlation > 0.8 for $\alpha \simeq 0.1$). At this point ($\alpha \simeq 0.1$), $\widetilde{T_1}$ is still nearly completely random: its correlation with $T_1$ is around $0.12$. Nevertheless, for larger training set sizes, this correlation significantly increases and reaches almost unity for $\alpha \simeq 0.5$. From the comparison between $I^{\text{test}}$ and $I^{\text{out}}$, we show that this method is equivalent to performing the correlations between ground truths and predicted matrices ($T_1$ and $\widetilde{T_1}$, $T_2$ and $\widetilde{T_2}$). The use of training and testing sets allows us to implement the two-layer neural network method on experimental data since we can verify the reconstruction of the matrices without knowing the ground truth for $T_1$ and $T_2$.

\subsubsection{Continuous object}
\begin{figure}[!htbp]
\centering
\includegraphics[width=0.7\textwidth]{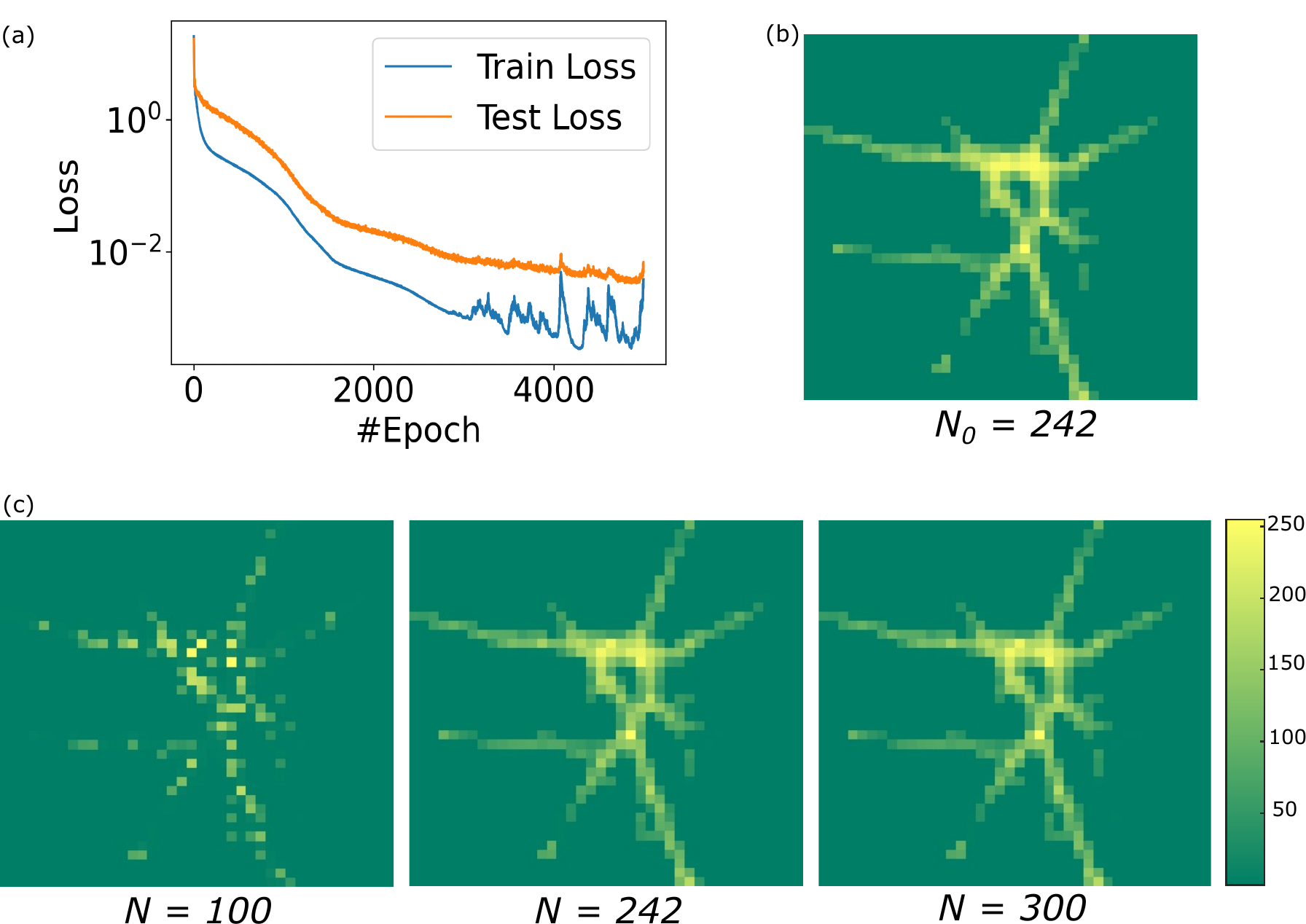}
\caption{\textbf{Continuous fluorescent object reconstruction by the 2-layer neural network.} (a) Example of train and test losses during the neural network training ($N = $300 case). (b) Ground truth image, simulated, $N_0$ is the number of targets. (c) Focus-combined images under different middle layer size $N$ (i.e rank) using phase conjugation of the retrieved $T_1$ from the 2-layer neural network showing an accurate prior information on the middle layer is not needed for TM retrieval in continuous sample. }
\label{continuous}
\end{figure}

By increasing the dimension of the hidden layer, our physics-based neural network is able to reconstruct more complex continuous objects. We first study the case of continuous objects through simulations. To simulate the ground truth of the fluorescence imaging process, we use the code from \cite{GithubNMF}. It is simulated by an i.i.d. Gaussian complex with tunable speckle grain size (adjusted in the Fourier plane thanks to a pupil function). In this way, we can control the speckle grain size by varying the pupil size of the pupil function, and we found that the speckle grain size does not have an impact on the TM retrieval. (In the simulation in Figure \ref{continuous}, we show a speckle grain size of 1 pixel and in Fig. \ref{shg}, we show the case of larger speckle grain of 4 pixels). To construct $T_1$, $T$ is further multiplied element-wise (Hadamard product) with the object vector, so that $T_1\in \mathbb{C}^{N_{\text{SLM}}\times N_{\text{target}}}$. No correlations are embedded in $T_1$ generation.
The intensity transmission matrix $T_2$ is simulated in a similar way, but by using real positive coefficients, corresponding to the intensity speckle patterns, $T_2\in \mathbb{R}^{N_{\text{target}}\times N_{\text{CAM}}}_{+}$. 

Here in our simulation, we chose as fluorescent continuous object a neuron with its connecting dendrites. We study the effect of the rank of the neural network (by changing the dimension of the hidden layer) on the final reconstruction. The training dataset is generated by randomly producing input patterns, and computing the corresponding output speckle images. These speckle images are obtained by propagating the input field with a field transmission matrix, and then multiply it with an intensity transmission matrix to simulate the final set of output fluorescence images. Once the neural network is trained, we phase conjugate $T_1$ and scan through each column to focus the images through the scattering media at all possible locations, and observe the images as they would appear at the object plane. Here we present the original target image without scattering and the focus combined image with phase conjugation $\widetilde{T_1}$, retrieved by the neural network \cite{Yang2019conjugphase}. Figure \ref{continuous} shows the combined foci with the shape of the targeted neuron. Although the rank (i.e. the dimension of the hidden layer) is a required input parameter, it can be approximated, as it will impact the intensity distribution but not the overall shape. Note that there is no condition on the sparsity of the object in this section.

\subsection{Experiment}

\subsubsection{Discrete object}
\begin{figure}[!htbp]
\centering
\includegraphics[width=0.8\textwidth]{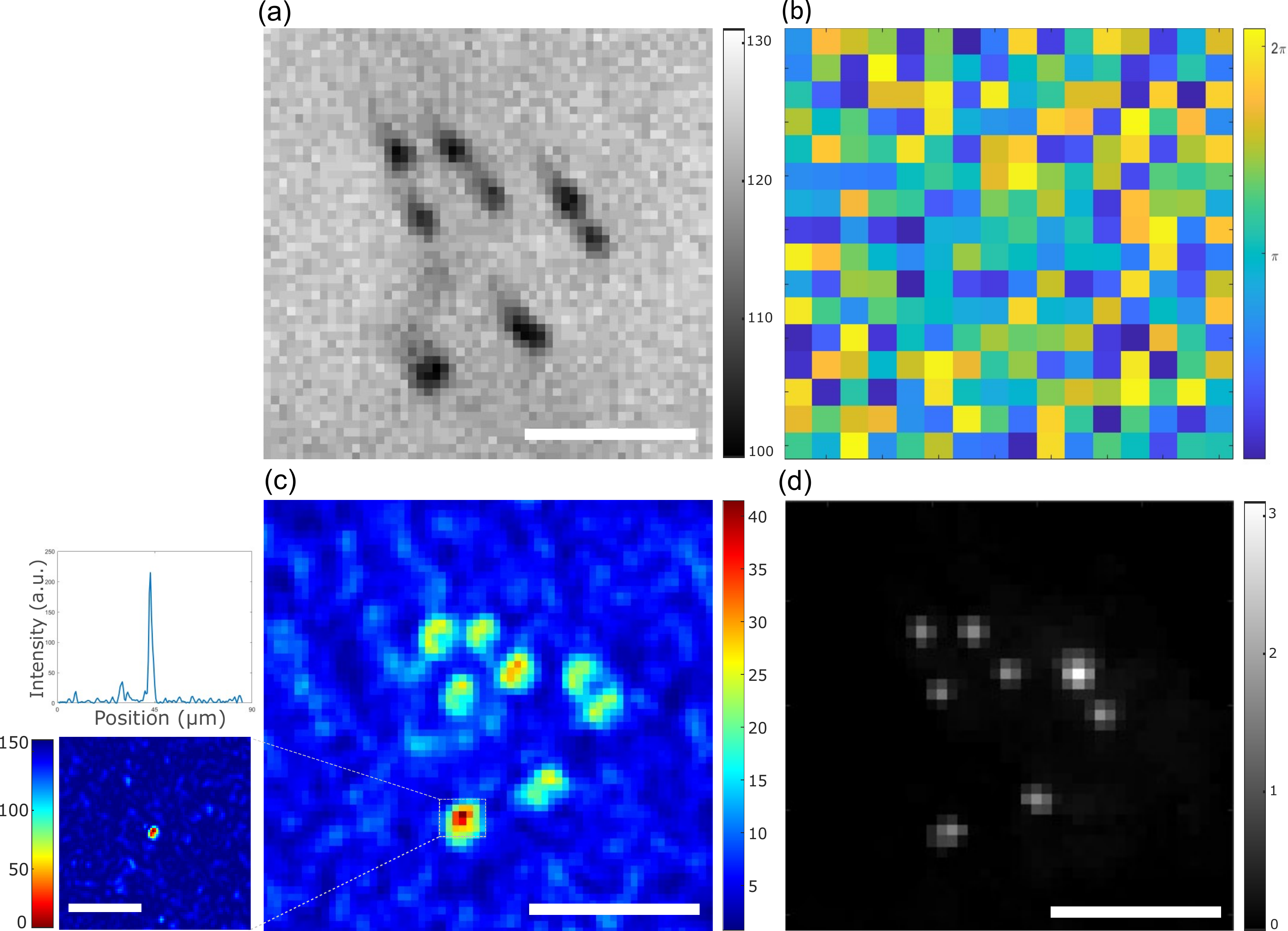}
\caption{\textbf{Experimental reconstruction of the fluorescent beads sample in epi-detection}. The rank of the hidden layer in the algorithm is $N_{\text{target}} = 9$, to ensure the reconstruction of all the beads. The number of patterns send onto the SLM is $N_{\text{pat}} = 15360$. (a) Ground truth, 8 beads, brightfield illumination, in the control camera. (b) Example of the SLM phase pattern projected to focus on one bead, found by phase conjugation of $\widetilde{T_1}$. (c) Sum of the images of the focus on each bead in the control camera. From each focus, the SNR is computed, and only the ones with an SNR higher than 20 are added to the sum. On the left, one focus with its associated intensity profile. (d) Reconstruction of the object, through scattering medium, from $T_2$ and using the cross-correlation procedure shown in \cite{boniface2020noninvasive}. The white bar corresponds to $10$ \textmu m.}
\label{exp_result}
\end{figure}

Experimentally, the measurement was first applied to 8 fluorescent beads of diameter $1$~\textmu m, placed on a holographic diffuser (Newport, 10DKIT-C1, 10$^o$ as the diffusion angle). A set of known input random patterns ($N_{\text{pat}} = 15360$) are displayed onto the SLM and the corresponding fluorescence speckles are recorded. With this input/output training set, the loss function of the 2-layer neural network is minimized and two transmission matrices $T_1$ and $T_2$ are finally obtained. In order to confirm the quality of this retrieval method, light is focused onto each bead using phase conjugation of $T_1$ \cite{vellekoop2012digital,Yang2019conjugphase}. When light is successfully focused, it indicates that both transmission matrices were well reconstructed. In Fig. \ref{exp_result}, one can see the sum of all the control camera snapshots (placed in transmission) of all the beads foci after phase conjugating $T_1$, and the non-invasive reconstruction of the object using $T_2$. From $T_2$ transmission matrix exploitation and the optical memory effect (OME), one can reconstruct the object shape \cite{boniface2020noninvasive,Zhu2021noninv,Feng1988CorrelationsAF}. The optical memory effect is a type of wave correlation that is observed in coherent fields, allowing control over scattered light through thin and diffusive materials \cite{Osnabrugge:17}.
Here only $16\times16$ macropixels of the SLM are modulated for the input, so our focusing enhancement will be impacted by the limited pixel counts. We expect to have an even higher signal to background ratio (SBR) with an increased number of SLM pixels. 

\subsubsection{Continuous object}
\begin{figure}[!htbp]
\centering
\includegraphics[width=0.85\textwidth]{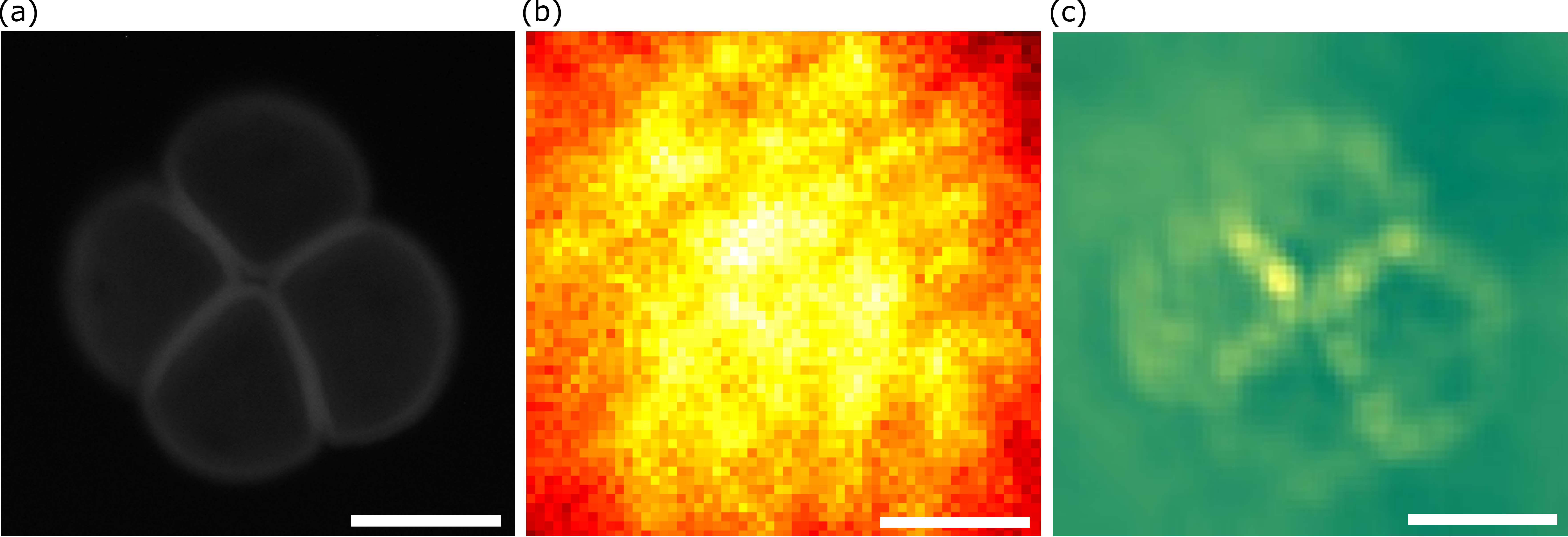}
\caption{\textbf{Experimental reconstruction of a fluorescent biological object in epi-detection.} The rank in the algorithm is set to $N_{\text{target}} = 53$. The number of patterns send onto the SLM is $N_{\text{pat}} = 5120$. (a) Ground truth, pollen seed image in transmission without scattering medium. (b) Example of the speckle we can get in epi-detection for an arbitrary SLM pattern with the pollen seed as the fluorescent object. (c) Reconstruction of the object, through the scattering medium, from pairwise deconvolution over the fingerprints of $T_1$. Scalebar is $10$~\textmu m.}
\label{pollen}
\end{figure}

The measurement was also applied to continuous objects like pollen seeds to show that the method can be used with 3D continuous objects and biological ones. The whole approach of acquisition, measurement, and processing is similar to beads samples: $P$ random incident wavefronts are produced by the SLM and sent onto the scattering medium, $P$ fluorescent speckles are respectively measured on the camera; the neural network is used to retrieve transmission matrices, $T_1$ is used to focus onto the pollen seed and the \textit{eigen patterns} are used to reconstruct the object thanks to OME. The dimension of the hidden layer is increased to match the number of equivalent discrete targets. The latter can be estimated from the minimization of the Frobenius norm of the Non Negative Matrix Factorization (NMF) residual is done \cite{boniface2020noninvasive, Zhu2021noninv}. 
To make this reconstruction work, we initialized the weight of the second layer with the result of the NMF algorithm over the output dataset \cite{Wang2013NMF}, in order to help the training of the neural network. This will help the convergence ratio of the neural network training.
Once $T_1$ and $T_2$ are retrieved, the fingerprint of each target is computed by phase conjugation over $T_1$ and a pairwise deconvolution is applied to reconstruct the object non-invasively \cite{Zhu2021noninv}.
Figure \ref{pollen} shows the ground truth image of a pollen seed and the neural network's reconstruction. 
The dataset used is the same as in \cite{boniface2020noninvasive}. 

This shows that the structure of the 2-layer neural network can also be used to retrieve a continuous object hidden in scattering medium in experiments since it can accept some priors such as the NMF over the output data to initialize the second layer.

\subsection{Other contrast mechanisms}
\begin{figure}[!ht]
\centering
\includegraphics[width=1.0\textwidth]{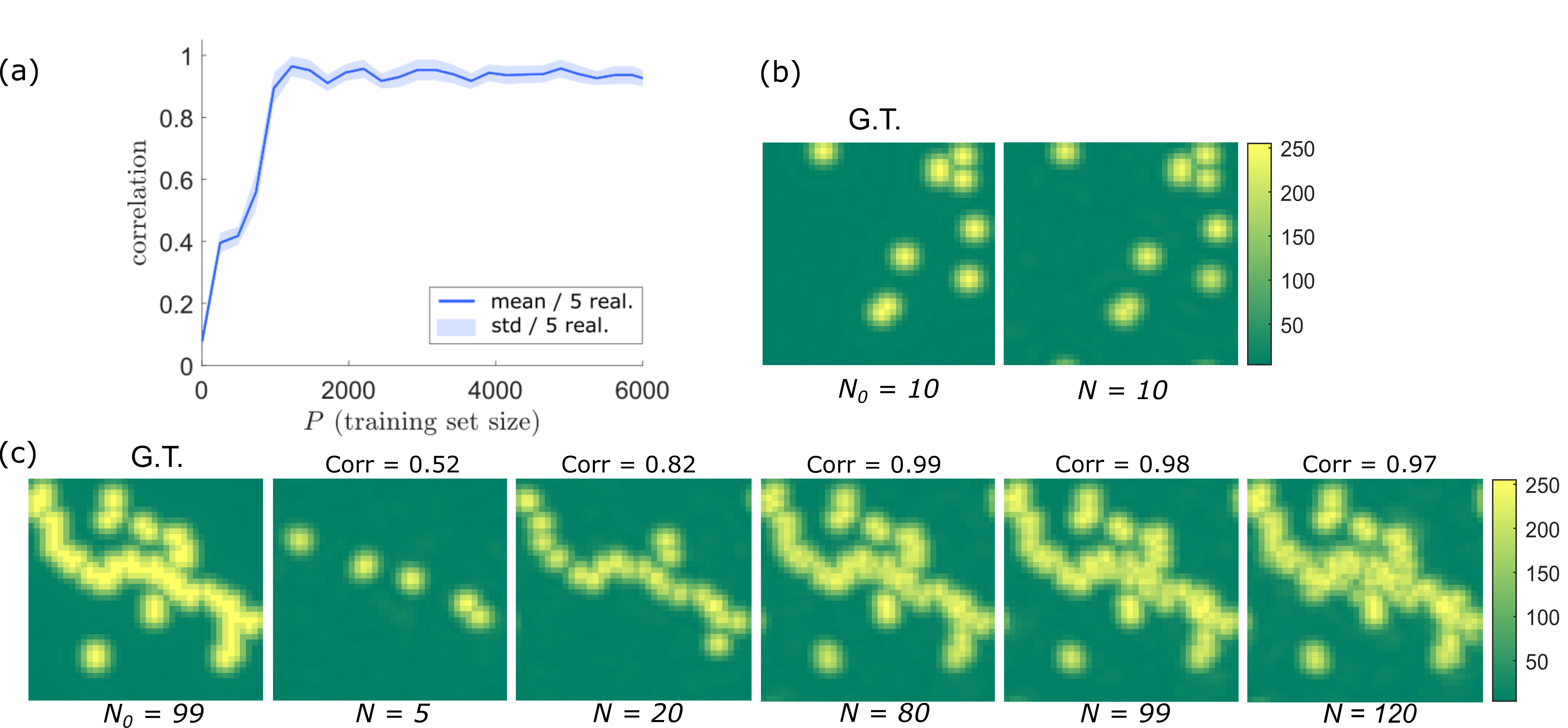}
\caption{\textbf{Simulation on the 2-layer neural network for SHG imaging process.} (a) Correlation curve between generated images and ground truth according to the training set size, over 5 realisations. At each realisation, the initialization of TMs is set to random matrices. 200 examples used in the test set. (b) Ground truth (G.T.) and focus-combined image, using phase conjugation of the retrieved $\widetilde{T_1}$ from the 2-layer neural network for a simulated discrete object (like beads). (c) Ground truth and focus-combined images plus their correlation with ground truth under different middle layer size $N$ using phase conjugation of the retrieved $\widetilde{T_1}$ from the 2-layer neural network for a simulated continuous object (like collagen tissue fibers).}
\label{shg}
\end{figure}

To illustrate the versatility of this neural network approach, simulations for another contrast mechanisms were performed. Here, we study numerically the case of imaging through a scattering medium with contrast from second harmonic generation (SHG).
Essentially, this is a coherent phenomenon which can be modelled by a pair of complex valued transmission matrices the same way as before, by simply changing the activation functions to $g_1 = .^2$ and $g_2 = |.|^2$. The forward model is then: 
\begin{equation}
    \begin{centering}
    x_2 = g_2(T_2g_1(T_1x_0)) = |T_2 (T_1 x_0)^2|^2
    \end{centering}
\end{equation}
The simulation procedure is the same as before, only the forward model has changed. It is more challenging to retrieve phase information correctly when compared to the fluorescent case. A correlation between the transmission matrices and the corresponding ground truth is shown in Fig.~\ref{shg}. Contrarily to previous results, the correlation does not reach exactly 1 and the training set size needed for high correlation is significantly higher than that in the case of fluorescence. This can be explained by the absence of a phase reference leaving some ambiguity for $T_2$. Despite this, the retrieved matrices should be sufficient for many applications, such as focusing, as can be seen on the simulation over a continuous object, using the same technique as described in the simulation section by simply changing the forward model over $T_2$. The focusing is again possible, as shown in Fig. \ref{shg}. As simulated objects, we chose polystyrene beads and an image of tissue collagen fiber, which are common SHG contrast sources.

\section{Discussion}
Our proposed 2-layer physics-based neural network approach enables the efficient retrieval of the two transmission matrices non-invasively, meaning without direct access to the object plane. With phase conjugation over $T_1$, we are able to focus and thanks to $T_2$ we can successfully reconstruct the object, hidden behind the scattering medium, provided there is some optical memory effect. In order to recover these two TMs, experimentally, for fluorescence, a significant amount of measurements is required: approximately 15000 input/output pairs of experimental data are used for training our network. 
Through numerical study, we confirmed that it is not necessary to have an accurate estimation of the rank of the object to have two transmission matrices retrieved. In the experiment, the limitation to the reconstruction of these complex objects (for example the pollen seeds) is the contrast of the measured speckle which decreases with a $1/\sqrt{N_{\text{target}}}$ dependency \cite{Boniface2019variance_optimization}. 
In terms of transmission matrix reconstruction, our approach is limited by the simple architecture of the network, because of few freedom degrees. Even so, this still allows us to add physical priors, such as changing the initialization and the nature of the components of the two TMs as well as the activation functions. For example, using the result of the non-negative matrix factorization \cite{Wang2013NMF} over $I^{\text{out}}$ to initialize $T_2$ helps find more targets than the previous 2-step approach based on NMF and phase retrieval \cite{boniface2020noninvasive} (see supplementary information - Influence of initialization). 
Finally, this approach is versatile and applicable to other contrast mechanisms, such as SHG where the positivity constraint on $T_2$ does not hold, but also Raman signal or 2-photons processes. The forward model is simply changed according to the physical phenomenon at stake, as shown in Table \ref{tableau}. 

\begin{table}[!ht]
    \centering
    \begin{tabular}{c|c|c}
    \hline
    Non-linearity & Coherent & Non-coherent \\
    \hline
    $n = 1$ & Raman: $g_1 = Id$, $g_2 = |.|^2$ & Linear Fluo.: $g_1 = |.|^2$, $g_2 = Id$\\
    \hline
    $n = 2$ & SHG: $g_1 = .^2$, $g_2 = |.|^2$ & 2-photon Fluo.: $g_1 = |.|^4$, $g_2 = Id$ \\
    \hline
    $n = 3$ & THG: $g_1 = .^3$, $g_2 = |.|^2$ & 3-photon Fluo.: $g_1 = |.|^6$, $g_2 = Id$\\
    \hline
    \end{tabular}
    \caption{Different phenomenon applicable to the 2-layer model with the right activation function.}
    \label{tableau}
\end{table}

\section{Conclusion}
Our study presented a physics-based machine learning method for characterizing light propagation through complex samples in order to focus on and reconstruct objects inside scattering media. Compared with other deep learning-based methods, our model is advantageous for data interpretation and incorporation of physical priors. Each node in the network has a physical meaning: it corresponds to the coefficients of the two transmission matrices. In order to test the quality of the neural network reconstruction, training and testing sets were used as is usually done in machine learning approaches. Compared to previous approaches regarding linear fluorescence \cite{boniface2020noninvasive}, this new one is more demanding in terms of measurements, because there are no priors on the algorithm. Furthermore, the phase information is not measured. However, the two-layer neural network is more general and less sensitive to noise (see supplementary information - Influence of noise). This approach can also be easily generalized and adapted to other contrast mechanisms such as 2-photon fluorescence or coherent processes where the previous method would not work. Moreover, the physics-based neural network approach is versatile, and additional physical priors may be added in the model to further enhance the capability of the method, such as memory effect \cite{freund1988memory} information or 3D composition of the sample.

\begin{backmatter}
\bmsection{Funding}
This research has been funded by the FET-Open (Dynamic-863203), European Research Council ERC Consolidator Grant (SMARTIES-724473, FunLearn-101020573), and Chan Zuckerberg Initiative (Deep Tissue Imaging grant no. 2020-225346). 

\bmsection{Acknowledgments}
The authors thank Dr. Lei Zhu and Dr. Fernando Soldevila for useful comments and technical support; Dr. Lorenzo Valzania and Dr. Hilton Barbosa de Aguiar for constructive discussions; Louis Delloye for valuable comments on this manuscript. 

\bmsection{Disclosures}
The authors declare no conflicts of interest.

\bmsection{Data Availability statement}
Data and codes underlying the results presented in this paper are available in Ref. \cite{Github}. 

\bmsection{Supplemental document}
See Supplement document for supporting content. 


\bibliography{sample}
\newpage

\end{backmatter}

\section*{Supplemental Information}

\subsection*{Experimental setup}
\begin{figure}[!ht]
\centering
\includegraphics[scale = 0.17]{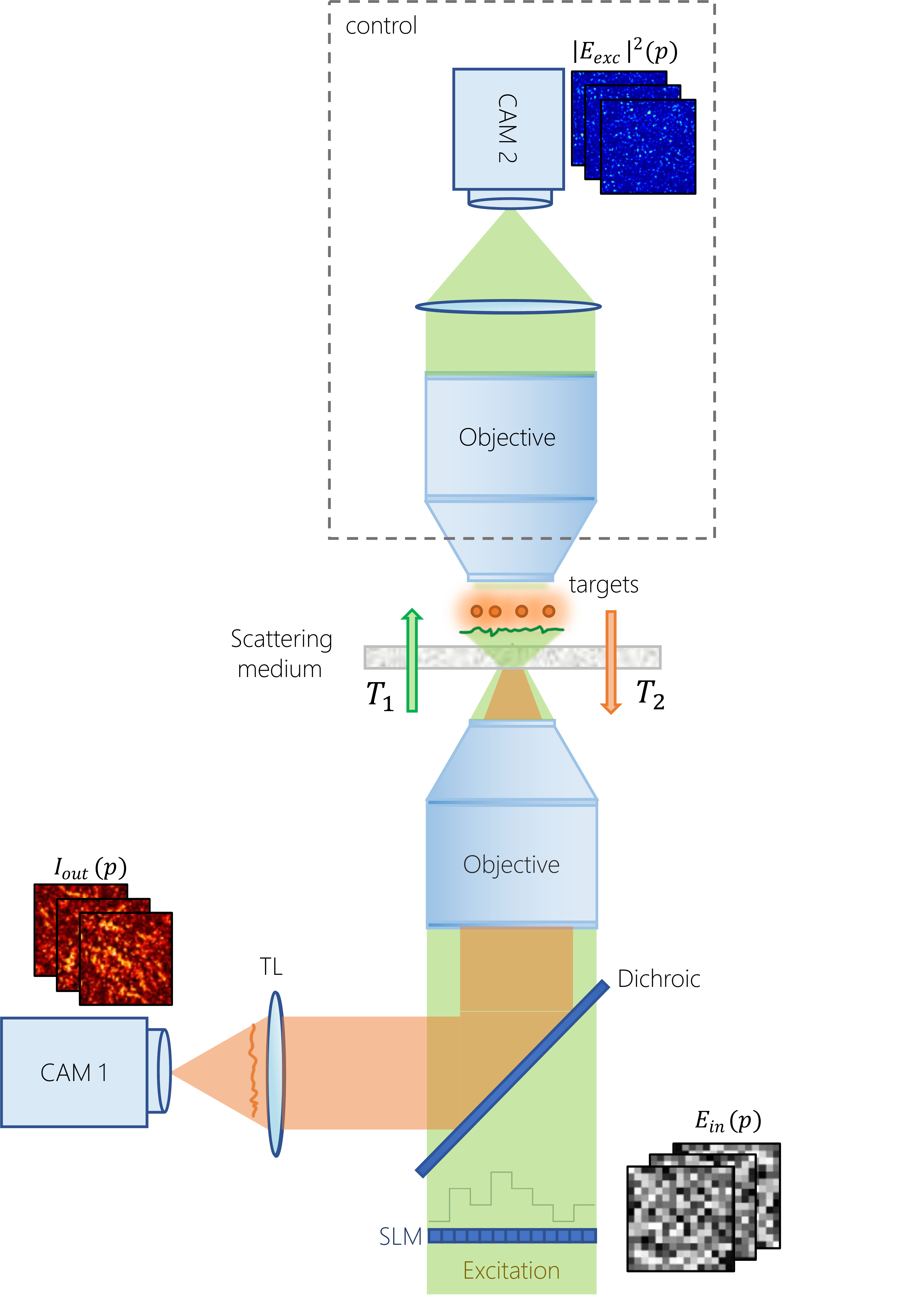}
\caption{\textbf{Schematic view of the complete experimental setup.} A randomly modulated speckle pattern, generated on the SLM, illuminates a fluorescent object (beads or pollen seeds), hidden behind the scattering medium. The fluorescent speckle is epi-detected on CAM 1. A second camera placed in transmission, CAM 2, monitors the illumination speckle in the plane of the beads and allows register a bright field image of the beads. CAM 2 is only used for passive monitoring. TL: tube lens.}
\label{experimental_setup}
\end{figure}

Full experimental setup with control camera on  Fig.~\ref{experimental_setup}. 

\subsection*{Rank: Hidden layer size}
\begin{figure}[!ht]
\centering
\includegraphics[scale = 0.23]{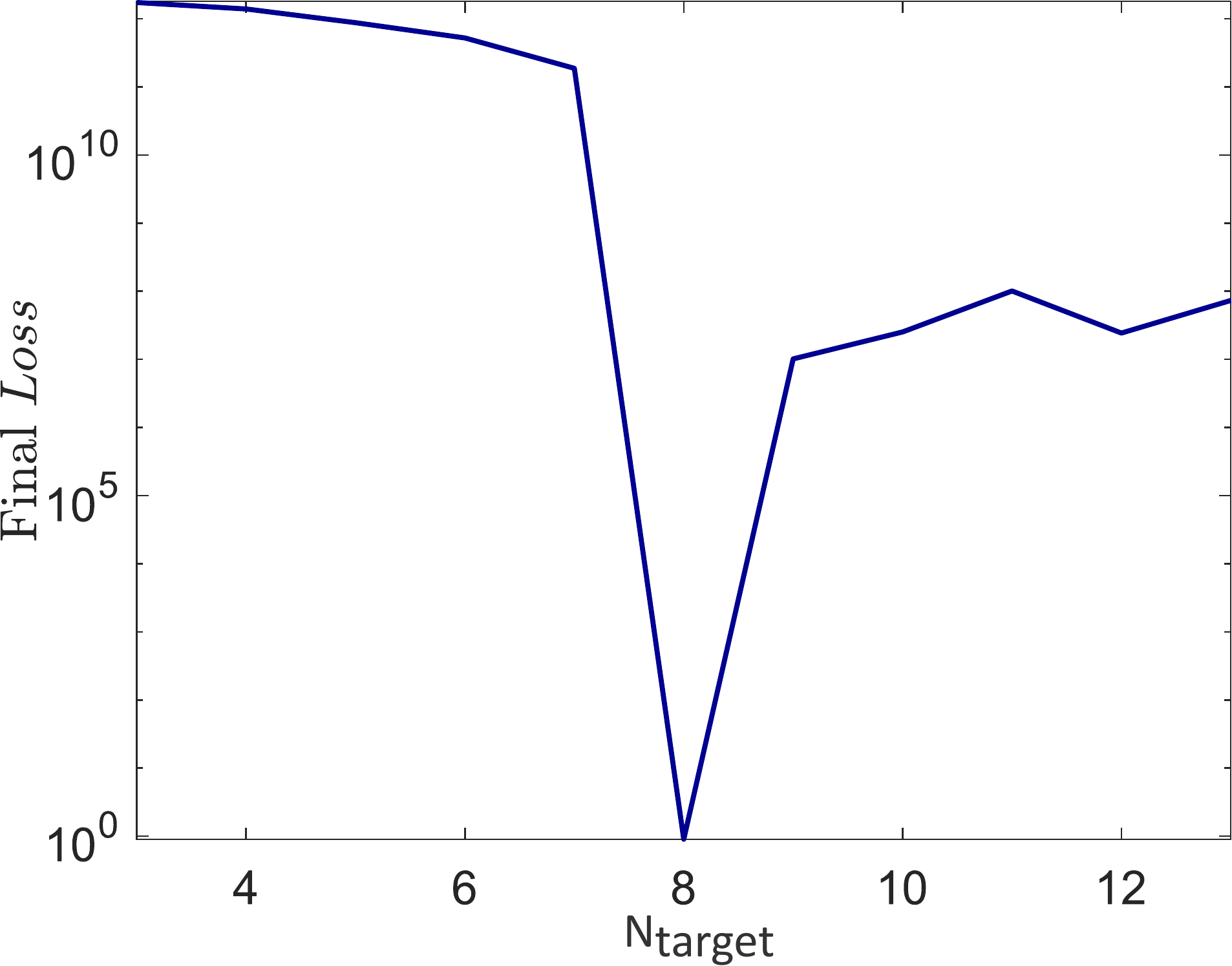}
\caption{\textbf{Dimension of the hidden layer.} Final Loss value (epoch= 1000) with respect to $N_{\text{target}}$. $N_{\text{target}} = 3, . . . , 13$ whereas the real dimension is $N_{\text{target}} = 8$.}
\label{rank}
\end{figure}

In our reflection geometry, we do not have access to the number of sources non-invasively easily. In order to find this number, we compute the loss function according to the hidden layer size and see which values helps to reach the minimum. As an example, we consider a fluorescent object constituted of N = 8 separate fluorescent targets and see what happens if this real number is under or over estimated. On  Fig.~\ref{rank}, one can see the minimum of the $Loss$ according to the rank value. 
When $N_{\text{target}} < 8$ or $N_{\text{target}} > 8 $, the minimum of the $Loss$ is far above 0. In the case $N_{\text{target}} < 8$, the network does not have a sufficient number of degrees of freedom to correctly fit the physical system. This prevents the network to generate patterns that faithfully reproduce the ground truths.
When the dimension of the hidden layer is correct, the final loss reaches a minimum. In that case, the number of neurons is high enough to reproduce the physical system with high fidelity. The Loss decreases almost to zero. 
When $N_{\text{target}} > 8 $, the $Loss$ is higher but lower than under 8. We think that these additional neurons do not play a major role in the fitting. It may be better to overestimate the dimension of the hidden layer and refine the matrices $T_1$ and $T_2$ of the network afterwards than the opposite. 

\subsection*{Influence of initialization} 
\begin{figure}[!ht]
\centering
\includegraphics[scale = 0.40]{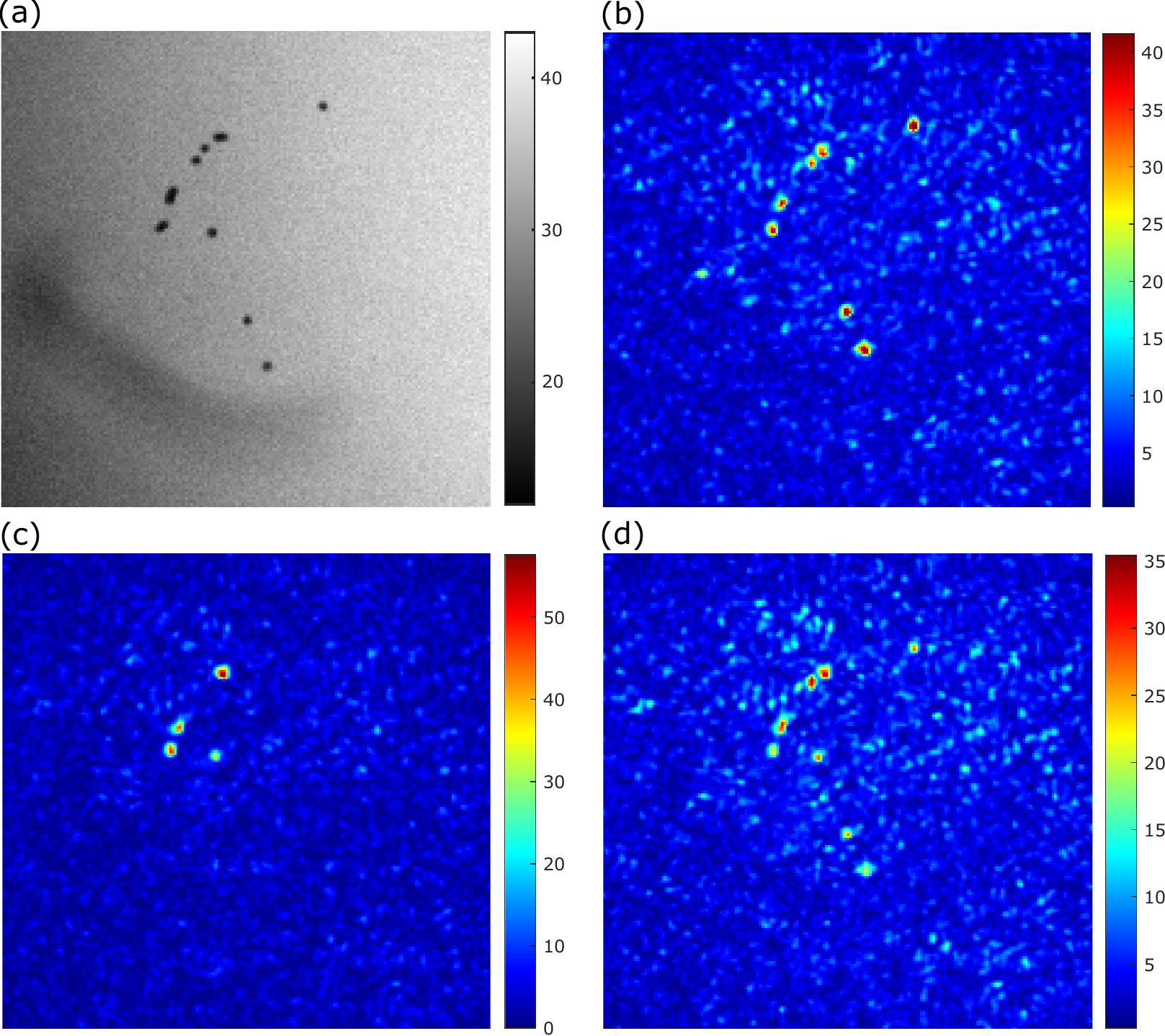}
\caption{\textbf{Influence of weight initialization in the second layer of the neural network.}(a) White light image in the control camera, ground truth of 9 fluorescent beads. (b) Sum of the focus on each bead in the control camera after phase conjugation thanks to $T_1$ found with NMF and Phase retrieval approach (\cite{boniface2020noninvasive}). $7/9$ beads found. Only the ones with an SNR higher than 20 are added to the sum. (c) Sum of the foci with $T_1$ from 2-layer neural network optimization with random initialization of the weights. $4/9$ beads found. Only the ones with an SNR higher than 20 are added to the sum. (d) Sum of the foci with $T_1$ from 2-layer neural network optimization with NMF initialization of the weights of the second layer. $8/9$ beads found.}
\label{NMF_supp}
\end{figure}

As said in the main text, the 2-layer neural network is versatile because it is simple to add a prior and give some physical information to it. For the second layer ($T_2$), there is a way to retrieve the weights: the Non-Negative Matrix Factorization (NMF) algorithm, from the speckles recorded on the epi-camera ($I^{\text{out}}$) \cite{Boniface2016TMPSF}. 
If the result of this algorithm is used as an initialization of the network weights, the reconstruction can be better, and faster because it helps the convergence of the gradient descent.
In the Fig.~\ref{NMF_supp}, we show on an experimental sample ($9$ fluorescent beads), how the weight initialization can improve the reconstruction. Using random initialization, it is possible to retrieve four of the nine beads. Initializing the 2-layer neural network with the result of the NMF approach, helps retrieving more beads, even more than the previous approach ($8/9$ vs $7/9$ with the NMF and Phase Retrieval algorithm). 

\subsection*{Influence of noise}
\begin{figure}[!ht]
\centering
\includegraphics[scale = 0.3]{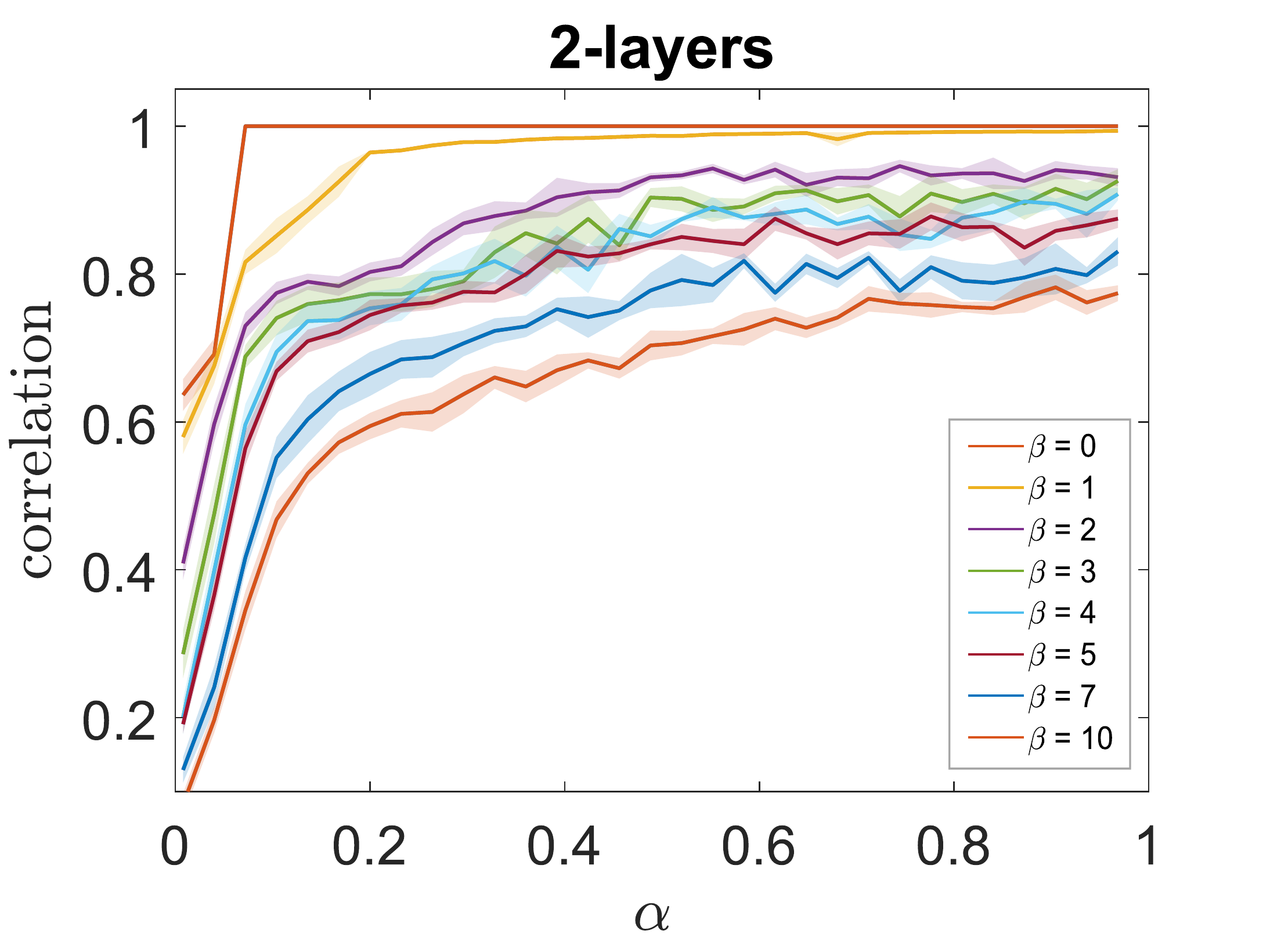}
\caption{\textbf{Influence of presence of noise in the data.} Correlations between $I^out$ and $I^{\text{test}}$, according to $\alpha = P/N_{\text{pat}}$, with P the training set size. The neural network under study has the following dimensions: $N_{\text{SLM}} = 64$, $N_{\text{target}} = 4$ and $N_{\text{CAM}}=64$. $\beta$ represents the level of noise, from 0 to 10 times the meaningful data. }
\label{noise}
\end{figure}

We investigated the versatility of the 2-layer neural network by looking at random noise added on top of the output data such as : 
\begin{equation}
    \begin{centering}
    I^{\text{out}} = T_{2}|T_{1}E^{\text{in}}|^2 + \beta  \langle T_{2}|T_{1}E^{\text{in}}|^2 \rangle R
    \end{centering}
\end{equation}
where $\beta$ controls the level of noise ($\beta = 0$, there is no noise, with $\beta = 1$, the amount of noise and meaningful signal are on average equal), and $R$ is the noise matrix (Gaussian distributed). 
The noise is not taken into account in the forward model, to mimic experimental results. 
On Fig.~\ref{noise}, one can see the results that show the robustness of the neural network to noise. The versatility of the two-layer neural network allows to re-find all the weights such that it minimizes as much as possible the $Loss$ function and fits the most correctly the physical system.

\end{document}